# Carbon *p* Electron Ferromagnetism in Silicon Carbide

Yutian Wang[1,5], Yu Liu[1,2,*], Gang Wang[2], Wolfgang Anwand[3], Catherine A. Jenkins[4], Elke Arenholz[4], Frans Munnik[1], Ovidiu D. Gordan[6], Georgeta Salvan[6], Dietrich R. T. Zahn[6], Xiaolong Chen[2], Sibylle Gemming[1,6], Manfred Helm[1,5], Shengqiang Zhou[1,†]

1. Institute of Ion Beam Physics and Materials Research, Helmholtz-Zentrum Dresden-Rossendorf, Bautzner Landstr. 400, 01328 Dresden, Germany
2. Research & Development Center for Functional Crystals, Beijing National Laboratory for Condensed Matter Physics, Institute of Physics, Chinese Academy of Sciences, Beijing 100190, China
3. Institute of Radiation Physics, Helmholtz-Zentrum Dresden-Rossendorf, Bautzner Landstr. 400, 01328 Dresden, Germany
4. Advanced Light Source, Lawrence Berkeley National Laboratory, Berkeley, California 94720, USA
5. Technische Universität Dresden, 01062 Dresden, Germany
6. Institute of Physics, Technische Universität Chemnitz , 09107 Chemnitz, Germany

*y.liu@hzdr.de

†s.zhou@hzdr.de

Ferromagnetism can occur in wide-band gap semiconductors as well as in carbon-based materials when specific defects are introduced. It is thus desirable to establish a direct relation between the defects and the resulting ferromagnetism. Here, we contribute to revealing the origin of defect-induced ferromagnetism using SiC as a prototypical example. We show that the long-range ferromagnetic coupling can be attributed to the *p* electrons of the nearest-neighbor carbon atoms around the $V_{Si}V_C$ divacancies. Thus, the ferromagnetism is traced down to its microscopic, electronic origin.

Unexpected ferromagnetism has been observed or theoretically predicted for numerous defective carbon based materials and wide-band gap semiconductors such as highly oriented pyrolytic graphite (HOPG), graphene, oxides and SiC[1-12], which provides an alternative for organic and semiconductor spintronics. As the origin of the ferromagnetism is different from that in conventional *d*-electron ferromagnets, any experimental evidence to reveal its origin will be crucial. Červenka *et al.*[13] demonstrated direct evidence that localized electron states at grain boundaries were one of the origins to induce ferromagnetism in HOPG. Ohldag *et al.*[14] proved that the ferromagnetism found in graphite originates from carbon π-states and hydrogen-mediated electronic states. Ugeda *et al.*[15] explained the formation of local magnetic moments by single vacancies in graphite. Recently, defect-induced ferromagnetism was found in SiC[8, 16-18]. Divacancies ($V_{Si}V_C$) are proven to exist in neutron irradiated and neon implanted SiC[8, 17]. Thus a question arises whether it is possible to establish a one-to-one correlation between the local moments and the specific orbitals/electrons in SiC.

On the other hand, SiC has been well known as a kind of economical and practical abrasive and a semiconductor for its application in high-temperature and high-voltage semiconductor electronics. As to our work, the good crystalline quality and the low concentration of impurities (please compare the relevant data in Ref. [8, 19, 20]) can remove the concerns whether the observed ferromagnetism originates from extrinsic factors (*e.g.* magnetic contamination, see Ref. [21, 22]). Moreover, the dynamics of defects and their charge states in SiC upon ion irradiation can be obtained by *ab initio* molecular dynamics simulations[23], rendering SiC an ideal testbed for the investigation of defect-induced ferromagnetism. Recent studies reveal that SiC could be a material with potential for applications in quantum optics and quantum information[24-27]. Therefore, direct experimental evidence for defect-induced ferromagnetism in SiC will have significant impact on other scientific areas related to defects.

In this paper, 6H-SiC single crystals irradiated with xenon ions are investigated to reveal the origin of its ferromagnetism. We present the results of X-ray absorption near-edge structure (XANES) and X-ray magnetic circular dichroism (XMCD) experiments at both the silicon and carbon K-edges in conjunction with sensitive magnetization measurements and first-principles calculations. These results show that the *p* electrons of the nearest-neighbor carbon atoms of $V_{Si}V_C$ are mainly responsible for the long-range ferromagnetic coupling. Our results provide important evidence for the origin of defect-induced ferromagnetism in SiC.

**Results**

**Magnetization measurements and sample selection.** As a prerequisite step, the pristine SiC wafer was checked for trace elements by using particle induced X-ray emission. The amount of transition metal impurities (Fe, Co and Ni) proves to be below the detection limit of around 1 µg/g (result shown in Fig. S1 in the supplementary material). Figure 1(a) exhibits the hysteresis loops of all implanted samples after subtraction of the diamagnetic background. The inset of Fig. 1(a) shows magnetization vs. field for sample 5E12 and the pristine SiC measured at 5 K. The pristine SiC is primarily diamagnetic with a weak paramagnetic contribution (see Figs. S2-S4 in the supplementary material for details). As shown in Fig. 1(a), SiC becomes ferromagnetic upon Xe ion irradiation. The strongest magnetization occurs for the sample 5E12, which is the sample subjected to the lowest fluence and with the least damage to the crystallinity (refer to Fig. S5). With rising fluence, the saturation magnetization ($M_s$) decreases from 0.72 $\mu_B$/vacancy to around 0.02 $\mu_B$/vacancy. The decrease of $M_s$ at large defect concentrations has also been observed in proton irradiated graphite[2, 28]. This is very probably due to damage to the crystalline order or due to the unfavorable spin-polarization when the defects are too close to each other[29]. The hysteresis loops measured for the sample 5E12 at 5 K and 300 K after subtracting the magnetic background from the pristine sample are shown in Fig. 1(b), indicating $M_s$ at 300 K is still around half of $M_s$ at 5 K and the transition temperature is higher than 300 K. Therefore, we focus on the sample 5E12 in the following investigation.

**Direct evidence for the origin of magnetism.** XMCD spectroscopy as an element-specific technique has been used to measure the magnetic contribution from different elements with partially occupied $3d$ or $4f$ subshells[30, 31]. Ohldag *et al.*[28] successfully applied this technique to investigate the magnetism at the carbon K-edge in proton irradiated HOPG. As it is possible to investigate the bonding state in SiC single crystals using XANES spectroscopy[32, 33], it is also possible to explore the magnetic contribution in defect-induced ferromagnetism in SiC with soft X-ray spectroscopy. Figure 2(a) shows the XANES spectra of the silicon K-edge for selected samples to investigate the source of the observed ferromagnetism.

Comparing with the pristine sample, the peak positions of samples after implantation are not changed, but the relative strength of the peak at 1848 eV decreases, which suggests an increase of defect density[34]. As shown in Fig. 2(b), the strength of the XMCD signal at the silicon K-edge is below the detection noise level in both the pristine sample and the sample 5E12. We may conclude that no spin-polarized states of 3$p$ electrons occur at silicon atoms, and thus silicon centers do not contribute to the ferromagnetism observed in the sample 5E12. Figure 2(c) shows the XANES spectra at the carbon K-edge of the sample 5E12 and the pristine sample measured at 300 K. Resonances around 285 eV and 290 eV correspond to the transition of carbon 1$s$ core-level electrons to $\pi^*$ and $\sigma^*$ bands, respectively[14, 28]. The resonance at 285 eV of the sample 5E12 is sharper than that of the pristine sample, indicating that the orbital hybridization at carbon is modified from the diamond-like $sp^3$-type carbon in pure SiC towards a more planar, graphitic $sp^2$-type carbon center, which leaves the orthogonal $p_z$ orbital unchanged and gives rise to the peak of $\pi^*$ bands[32]. This reflects a change of the local coordination from the tetrahedrally coordinated carbon atom in pristine SiC to the three-fold bound carbon site. In sharp contrast to the silicon K-edge, a clear XMCD signal appears at the carbon K-edge as shown in Fig. 2(d). Therefore, the defect-induced ferromagnetism originates from a spin-polarized partial occupancy of the $p_z$ orbitals at carbon atoms close to defect sites in SiC. It is worth noting that an XMCD peak at around 280 eV (Fig. 3d) appears well below the onset of the $\pi^*$ resonance. This peak was also observed in graphite[14]. This intriguing feature is not yet fully understood.

**Discussion**

According to the results provided by positron annihilation spectroscopy (see Figure S6 in the supplementary material), divacancies $V_{Si}V_C$ are the dominating defect type in our samples. Note that the nearest-neighbor atoms of $V_{Si}V_C$ include three carbon atoms as well as three silicon atoms. Why is the magnetic signal observed only at the carbon sites? To answer this question, first-principles calculations were employed. As shown in Fig. 3(a), 90% of the spin polarization with a total moment of 2 $\mu_B$ due to one divacancy $V_{Si}V_C$ is contributed by the valence states of the carbon atoms. This explains why XMCD is only observable at the carbon K-edge. Furthermore, when comparing the partial spin-resolved DOS of nearest-neighbor carbon atoms with that of other carbon atoms, it is visible [see Figure 3(b)] that 85% of the magnetic moments originate from the three nearest-neighbor carbon atoms. In the Si-C system, as carbon has higher electronegativity than silicon, unpaired electrons around carbon atoms should be more localized than those around silicon. A Mulliken population analysis indicates that in the unperturbed SiC bulk the Si-C bonds are already polar in accordance with the respective electronegativities: Partial charges of -0.32 e on carbon atoms and of +0.32 e on silicon atoms are calculated for the pristine bulk at the Mulliken level. In the vicinity of the divacancy carbon atoms show a trend towards larger partial charges (-0.38 e), whereas the silicon partial charges close to the divacancy are nearly unchanged. Spin polarization thus mainly appears at those carbon atoms that are located around the divacancies. According to Fig. 3(c), our calculation indicates that most of the magnetic moments (90%) originate from the *p* states of nearest-neighbor carbon atoms of $V_{Si}V_C$. Due to the attraction of the remaining adjacent silicon atoms, the nearest-neighbor carbon atoms will slightly move away from the $V_{Si}V_C$. This structure change from the unperturbed four-fold bulk coordination to a more planar three-fold bound state is connected with s-p-rehybridization at the C atoms in the close vicinity of $V_{Si}V_C$. Concomitantly, this distortion will modify the electronic structure locally towards a higher degree of *$sp^2$* bonding orbitals and a singly occupied *p*-type lone pair at the C atoms. Thus those outermost orbitals will acquire significant π character and the magnetic moments are mainly contributed by *p* electrons, as shown in Fig. 3(d). This analysis corroborates our interpretation of the XMCD experiment: the XMCD signal of SiC after irradiation is thus assigned to *p* electrons.

In conclusion, in this work we investigated the magnetic properties of 6H-SiC after xenon irradiation. X-ray absorption spectroscopy at both the silicon and carbon K-edges combined with sensitive magnetization measurements and first-principles calculations are used to understand the origin of defect-induced ferromagnetism. The results give strong evidence that the *p* electrons of the nearest-neighbor carbon atoms of $V_{Si}V_C$ are mainly responsible for the observed ferromagnetism. These results provide valuable insight into comprehending the phenomena of defect-induced ferromagnetism in SiC, graphitic and other carbon-based materials, and will encourage the exploration of the origin of defect-induced ferromagnetism in other promising materials such as graphene and oxides.

**Methods**

**Sample preparation.** A commercial one-side-polished semi-insulating 6H-SiC (0001) single crystal wafer was cut into pieces for ion irradiation. The concentrations of transition metal impurities (Fe, Co and Ni) prove to be below the detection limit of particle induced X-ray emission (see Fig. S1 in the supplementary material). Four SiC pieces were implanted by xenon ions with fluence values of $5 \times 10^{12}$, $1 \times 10^{13}$, $5 \times 10^{13}$, $1 \times 10^{14}$ cm$^{-2}$ at an energy of 500 keV at room temperature, which were subsequently labelled as 5E12, 1E13, 5E13, and 1E14, respectively. During implantation, the samples were tilted by 7 degrees to reduce the channelling effect. The corresponding displacements per atom (DPA) values have been calculated by Stopping and Range of Ions in Matter (SRIM)[35] to be 0.023, 0.047, 0.23, and 0.47, respectively. The distribution of irradiation-induced damage predicted by SRIM for xenon ions is more uniform and closer to the surface than that produced by 140 keV neon ions[17].

**Measurements.** All samples were measured with a superconducting quantum interference device (SQUID-MPMS or SQUID-VSM, Quantum Design). The magnetization is determined according to the total vacancies calculated using SRIM[35]. Both XANES and XMCD spectroscopies at the silicon and carbon K-edges were obtained at the Advanced Light Source (Berkeley Lab). The spectra of the silicon

K-edge were measured at BL6.3.1 under a magnetic field of -2 and 2 T at 77 K, while the carbon K-edge spectra were measured at BL4.0.2 with the possibility of using a X-ray photon energy as low as 100 eV (note that the carbon K-edge is around 285 eV) and applying an external field of -0.5 and 0.5 T at 300 K. The typical spectral resolution for both beamlines is E/ΔE ~ 5000 (see Ref. 14). In the measurements, total electron yield (EY) mode is chosen, which usually collects the signal from the topmost 5-10 nm of the sample[36].

**Calculation parameters.** First-principles calculations were performed using the Cambridge Serial Total Energy Package[37]. Spin-polarized electronic structure calculations were performed using the Perdew-Burke-Ernzerhof functional[38] for the exchange-correlation potential based on the generalized gradient approximation. The core-valence interaction was described by ultrasoft pseudopotentials[39], and to represent the self-consistently treated valence electrons the cutoff energy of the plane-wave basis was set to 310 eV. We calculated the total spin-resolved density of states (DOS) and the partial spin-resolved DOS of silicon atoms and carbon atoms in a 4 × 4 × 1 *6H*-SiC supercell containing one axial $V_{Si}V_C$ [$Si_{95}(V_{Si})C_{95}(V_C)$]. The calculation presented in this paper is for neutral $V_{Si}V_C$. With the minimum distance between adjacent $V_{Si}V_C$ larger than 12 Å, this structure allows long-range ferromagnetic coupling[8]. The content of the spin polarization contribution is determined by comparing the integrated DOS below the Fermi level.

**Acknowledgements**

The work is financially supported by the Helmholtz-Gemeinschaft Deutscher Forschungszentren (VH-NG-713, VH-VI-442 and VH-PD-146). Y. Wang thanks the China Scholarship Council for supporting his stay at HZDR. The Advanced Light Source is supported by the U.S. Department of Energy under Contract No. DE-AC02-05CH11231. G. Wang, and X. L. Chen also thank the support by the National Natural Science Foundation of China (Grant Nos. 51322211, 90922037, 51072222 and 51272276).

**Author contributions**

S. Z. conceived the experiment. Y. W. prepared the samples and performed the measurements for magnetic properties. Y. L. and G. W. made the calculation. W. A. carried out the PAS experiment. F. M. did the PIXE measurement. C. J. and E. A. assistant the XAS experiment. O. G., G. S. and D. Z. performed the Raman measurements. S. G., X. C. and M. H. supervised the work. All authors have participated the manuscript preparation and discussion.

**Additional information**

**Competing financial interests:**

The authors declare no competing financial interests.

Figure captions

Figure 1 (a) Ferromagnetic hysteresis loops of samples 5E12, 1E13, 5E13, 1E14 at 5 K after subtracting the magnetic background from the pristine sample. The inset shows the as-measured magnetization vs. field of the sample 5E12 and the pristine sample at 5 K. (b) Hysteresis loops of the sample 5E12 at 5 K and 300 K.

Figure 2 X-ray absorption spectra measured in EY (electron yield) mode for the sample 5E12 and the pristine sample: (a) XANES of the silicon K-edge at 77 K, (b) XMCD at the silicon K-edge at 77 K, (c) XANES of the carbon K-edge at 300 K. (d) XMCD at the carbon K-edge at 300 K.

Figure 3 The electronic structure of $Si_{95}(V_{Si})C_{95}(V_C)$: (a) The total spin-resolved DOS and the partial spin-resolved DOS of silicon atoms and carbon atoms, respectively. (b) Comparison of partial spin-resolved DOS of nearest-neighbor carbon atoms and others. (c) Comparison of the partial spin-resolved DOS of *s* and *p* electrons of nearest-neighbor carbon atoms of $V_{Si}V_C$. (d) The structure and spin density isosurface (0.08 e/Å$^3$, in purple) around one of the nearest-neighbor carbon atoms. The carbon atom is in grey in the middle and silicon atoms are in yellow. The arrows indicate the bond angles of Si-C-Si. The dashed line and square indicate the direction and the location of the adjacent silicon vacancy part ($V_{Si}$) within $V_{Si}V_C$, respectively.

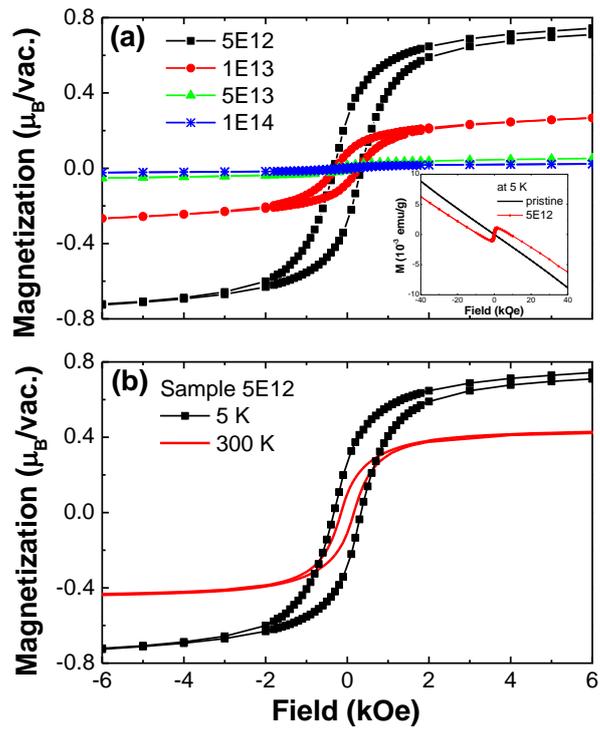

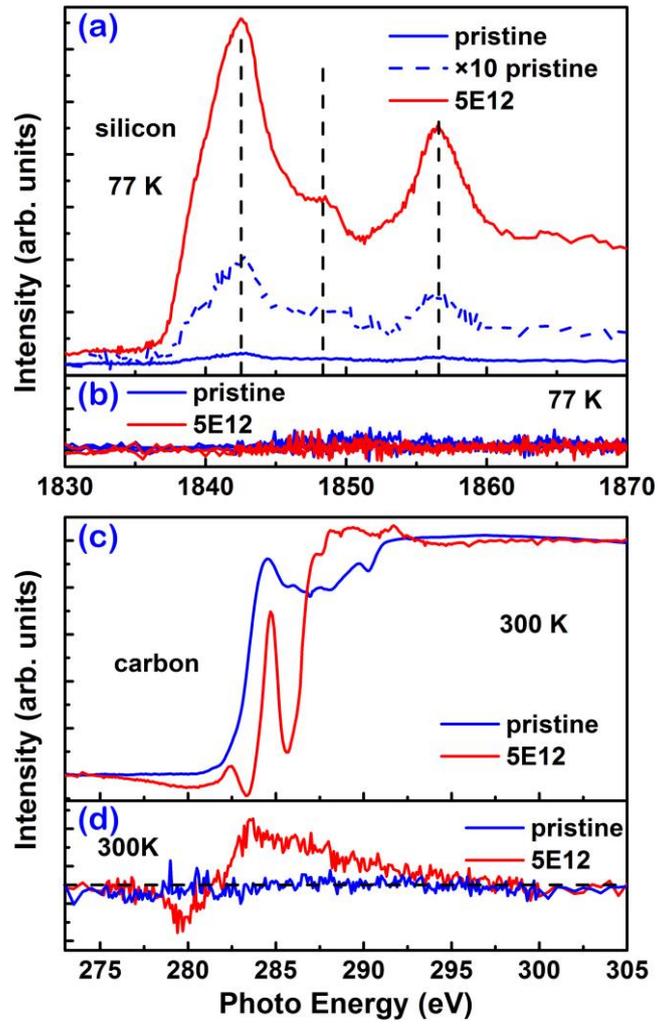

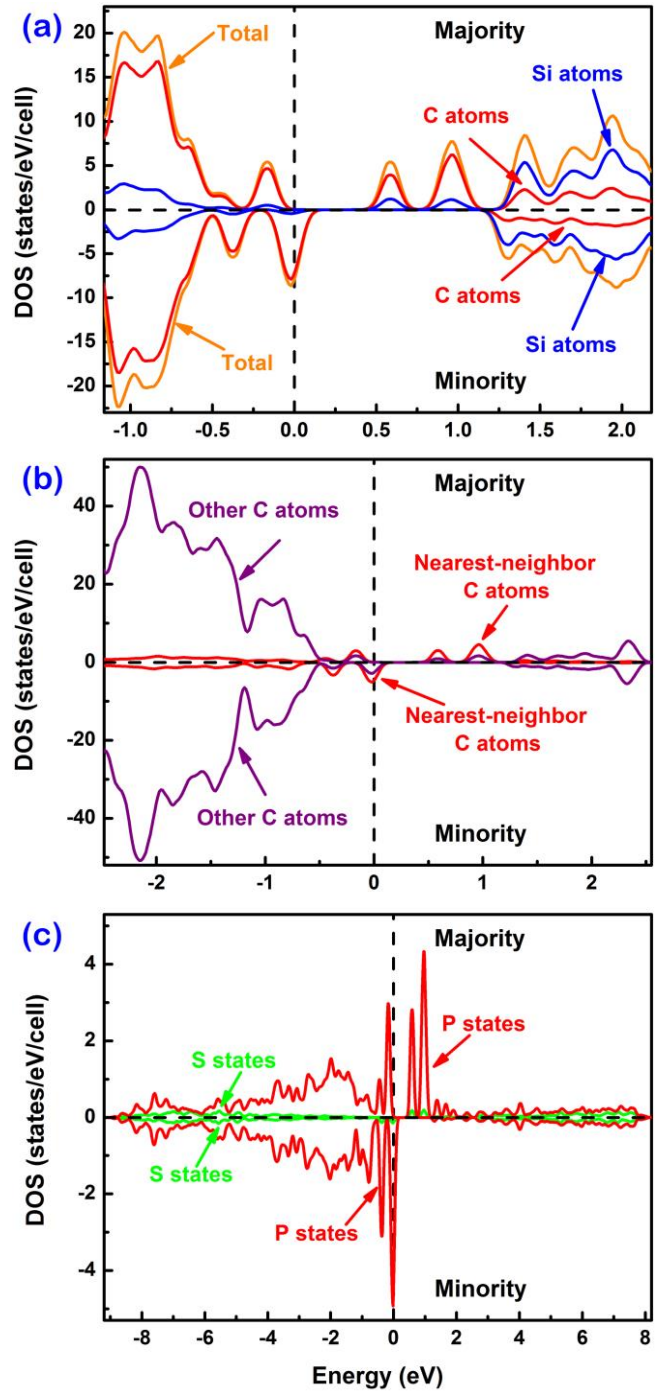
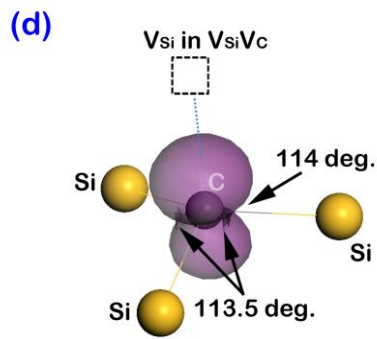

# Supplementary materials

**Section 1: Transition metal free SiC substrates**

To exclude the presence of possible transition metals (Fe, Co and Ni) in SiC substrates, we have performed Particle-induced X-ray emission (PIXE) using 2 MeV protons with a broad beam of 1 mm$^2$. The result is shown in Fig. S1. As stated in ref. 1, PIXE is a sensitive method to detect trace impurities in bulk volume without structural destruction. In the spectrum, the narrow peak is from Si K-line X-ray emission. The broad peak is due to the Secondary Electron Bremsstrahlung background. If there are any transition metal impurities, they are below the detection limit of around 1 µg/g.

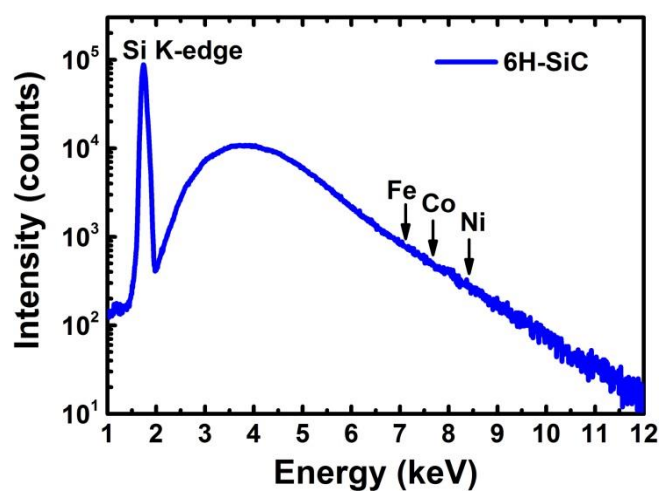

Figure S1: PIXE spectrum for the 6H-SiC wafer by a broad proton beam. Within the detection limit, no Fe, Co or Ni contamination is observed.

**Section 2: Magnetic properties of pristine SiC**

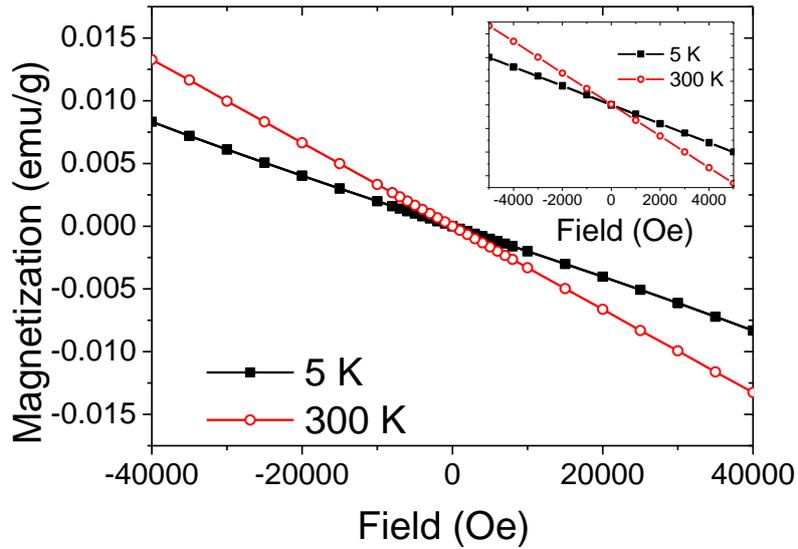

Figure S2: Field magnetization (open symbols) of pristine 6H-SiC measured at 5 K and 300 K. The sample is primarily diamagnetic. However, there is a difference in magnetic susceptibility at 5 K and 300 K. A slight deviation from the linear dependence of magnetization on field is observed at 5 K, which indicates a small paramagnetic contribution. The inset shows the zoom of the corresponding curves at the low field part. At both temperatures, the sample does not show ferromagnetic hysteresis.

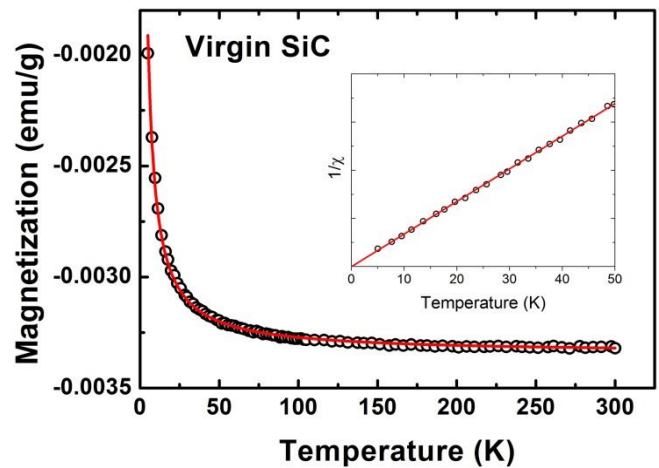

Figure S3: Temperature dependent magnetization (open symbols) of pristine 6H-SiC measured under a field of 10000 Oe. The result (open circles) can be fitted well (red solid line) by considering a paramagnetic contribution (Curie's law) with a diamagnetic background. The inset shows inverse

susceptibility (after subtracting the diamagnetic background) *vs.* temperature with a linear, purely paramagnetic behavior with no sign of magnetic ordering.

In this part, we present the magnetic properties of pristine SiC. Figure S3 shows the field dependent magnetization (open symbols) of pristine 6H-SiC measured at 5 K and 300 K. The sample is primarily diamagnetic. However, there is a difference in magnetic susceptibility at 5 K and 300 K. A slight deviation from the linear dependence of magnetization on field is observed at 5 K, which indicates a small paramagnetic contribution. The inset shows the zoom of the corresponding curves at the low field part: there is no ferromagnetic hysteresis. The weak paramagnetism in pristine SiC is due to the intrinsic defects, which have been identified by electron spin resonance spectroscopy [2, 3]. In the following, we further confirm that the substrate we used for this study contains ONLY spin ½ paramagnetic center by magnetization measurements, without any ferromagnetic inclusions as often observed in graphite [4]. Figure S4 shows the temperature dependent magnetization measured under a field of 10000 Oe. The large contribution represents the diamagnetic background, which is essentially temperature independent. We can well fit the curve by Curie's law:

$$M = M_0 + C\frac{B}{T} \qquad (1)$$

Where $M_0$ is the diamagnetic background, $B$ is the magnetic field, $T$ is the temperature and $C$ is the Curie constant. The inset shows the $1/M$ vs. temperature. The curve shows a linear behavior crossing the zero point, indicating the non-interacting nature between the paramagnetic centers.

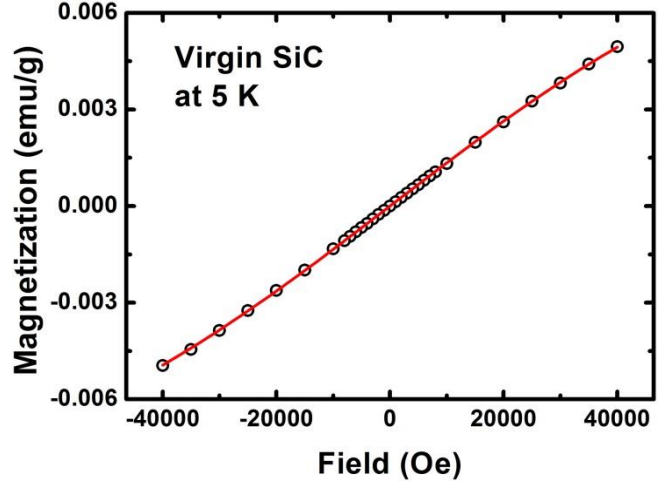

Figure S4: Fitting of the paramagnetism in the SiC substrate by Brillouin function with *J*=S=1/2.

The field-dependent magnetization of the pristine 6H-SiC was measured at 5 K. After subtracting the diamagnetic background, the result is shown in Fig. S5. No hysteresis loop is observed down to 5 K. The paramagnetism can be well fitted by a Brillouin function:

$$M = NJ\mu_B g_J \left[\frac{2J+1}{2J}\coth\left(\frac{2J+1}{2J}\alpha\right) - \frac{1}{2J}\coth\left(\frac{\alpha}{2J}\right)\right] \quad (2),$$

where $\alpha = g_J J \mu_B H / k_B T$, the $g_J$ factor is about 2 as obtained from electron spin resonance measurement [2, 3], $\mu_B$ is Bohr magneton, and *N* is the number of spins. The curve can only be well fitted by J = S = 0.5 corresponding to free electron spins. Thus in the pristine sample, the paramagnetic part is purely spin ½ paramagnetism, which is due to the intrinsic defects in SiC [2, 3].

**Section 3: UV Raman results: no graphite or graphene formation**

As ion irradiation induces damage only near the surface, UV-Raman spectroscopy with a He-Cd laser of 325 nm is used to characterize the structure after ion irradiation. Figure S2 exhibits UV-Raman spectra measured at room temperature for samples 5E12, 1E13, 5E13, 1E14 and the pristine sample, respectively. The corresponding phonon modes in 6H-SiC are identified [5] and no secondary phase is detected within the measurement sensitivity. The strength of the peaks decreases with increasing fluence, which indicates that the crystal structure is damaged upon ion bombardment. The samples 5E12 and 1E13 maintain relatively good crystallinity, while the structures of samples 5E13 and 1E14 suffer too much damage so that the sharp peaks vanish.

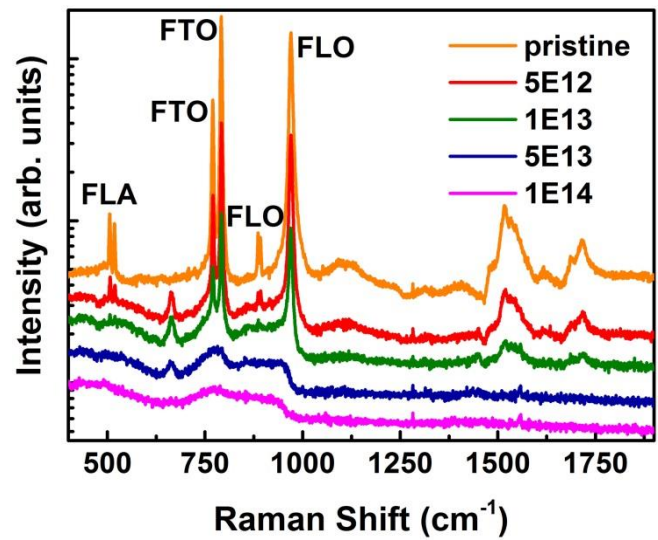

Figure S5: Room temperature UV-Raman spectra of samples 5E12, 1E13, 5E13, 1E14 as well as the pristine sample. The folded transverse acoustic (FTA) and optic (FTO) and longitudinal acoustic (FLA) and optic (FLO) modes in 6H-SiC are identified in these spectra. The peaks at higher wavenumbers are second order modes.

**Section 4: Divacancy identified in the ion implanted sample**

In order to clarify the nature of defects which were created by the Xe ion implantation, positron annihilation Doppler broadening spectroscopy (DBS) was applied. DBS is an excellent technique to detect open volume defects from clusters consisting of several vacancies down to a mono-vacancy. The positron in a crystal lattice is strongly subjected to repulsion from the positive atom core. Because of the locally reduced atomic density inside the open volume defects, with a lower local electron density, positrons have a high probability to be trapped and to annihilate with electrons in these defects by the emission of two 511 keV photons. Monitoring of the 511 keV annihilation radiation was performed by DBS. The Doppler broadening of the 511 keV annihilation line is mainly caused by the momentum of the electron due to the very low momentum of the thermalized positron. There is one main parameters, S (shape), obtained from the 511 keV annihilation line. The S parameter reflects the fraction of positrons, annihilating with electrons of low momentum (valance electrons). Therefore, the S parameter is mainly a measure for the open volume in the material. The S parameter is defined as the ratio of the counts from the central part of the annihilation peak (here 510.17 keV – 511.83 keV) to the total number of counts in the whole peak (498 keV – 524 keV). DBS measurements were carried out with the mono-energetic slow positron beam ''SPONSOR'' at HZDR [6] at which a variation of the positron energy E from 30 eV to 36 keV with a smallest step width of 50 eV, if required, is possible. The energy resolution of the Ge detector at 511 keV was (1.09 $\pm$ 0.01) keV, resulting in a high sensitivity to changes in material properties from surface to depth of several μm.

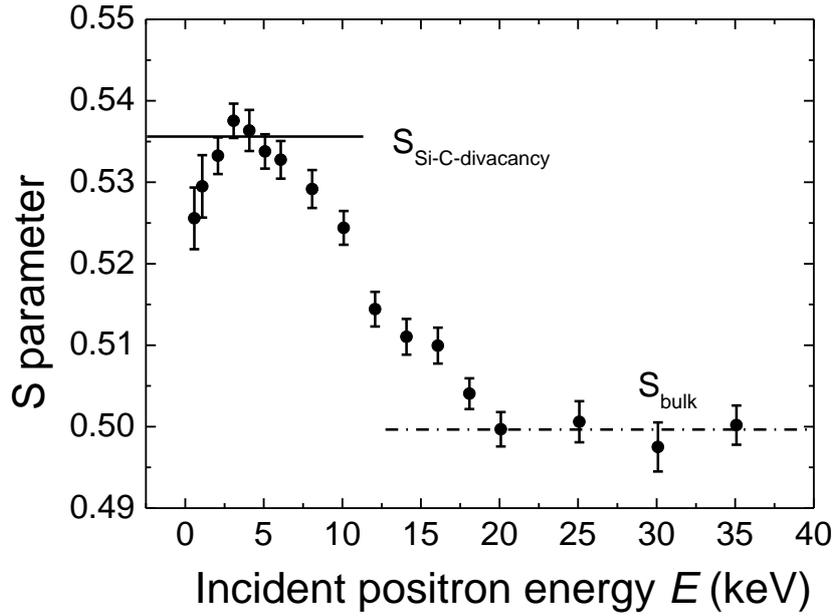

Figure S6: S parameter versus incident positron energy of the Xe implanted 6H-SiC sample.

Figure S6 shows the measured S parameter versus the incident positron energy. The maximum of the S parameter around positron energy of 3 – 4 keV corresponds to positron annihilation in the defects created by ion implantation within a depth of about 200 nm. For higher positron energies the positron annihilation is shifted more and more to the undamaged bulk material, the S parameter decreases and finally reaches the bulk value at 20 keV (dash-dot line in Fig. S6). The difference of the S values between the undamaged bulk $S_{bulk}$ and the defects $S_{defect}$ is a measure for type and concentration of these defects. Comprehensive investigations of defects in 6H-SiC were already done in the past. A relation between the S parameter and the number of agglomerated Si-C divacancies is published in [7] and shown as a scaling curve. From this scaling curve the S parameter of the Si-C divacancy was taken and plotted as a solid line in Fig. S6. It is clearly visible that the S parameter of the defects in the implanted range agrees well with the value of S for the Si-C divacancy which leads to the conclusion that Si-C divacancies were created by the Xe implantation into 6H-SiC.